
\font\subtit=cmr12
\font\name=cmr8

\input harvmac
\def\UWrLMU#1#2#3#4
{\TITLE{LMU-TPW \number\yearltd-#1}
{ITP UWr #2/\number\yearltd}{#3}{#4}}
\def\TITLE#1#2#3#4{\nopagenumbers\abstractfont\hsize=\hstitle\rightline{#1}
\vskip 1pt\rightline{#2}
\vskip 1in
\centerline{\subtit #3}
\vskip 1pt
\centerline{\subtit #4}\abstractfont\vskip .5in\pageno=0}%
\UWrLMU{20}{856}
{OPERATOR FORMALISM ON THE Z$_n$ SYMMETRIC ALGEBRAIC
CURVES}{}
\centerline{F. F{\name ERRARI}$^{a}$,
J. S{\name OBCZYK}$^b$, W. U{\name RBANIK}$^c$}\smallskip
$^a${\it Sektion Physik der Universit\"at M\"unchen,
Theresienstr. 37, 8000 M\"unchen 2, Fed. Rep. Germany}\smallskip
$^b${\it Institute for Theoretical Physics, Wroclaw University, pl.
Maxa Borna
9, 50205 Wroclaw, Poland}\smallskip
$^c${\it Academy of Economics, Wroclaw, Poland}
\vskip 2cm
\centerline{ABSTRACT}
{\narrower
In this paper it is shown that the $b-c$ systems
on a Riemann surface $\Sigma_g$ of genus $g$
and with abelian group of internal
automorphisms $Z_n$ are equivalent to multivalued field theories on the
complex plane {\bf C}.
To this purpose, the fields $b$ and $c$ are expanded using suitable
bases of tensors that are multivalued
on {\bf C} and singlevalued on $\Sigma_g$.
The amplitudes of the $b-c$ systems on $\Sigma_g$
are then recovered exploiting simple normal ordering rules on the
complex plane. Finally, we construct a conformal field theory in
{\bf C} having as primary operators twist fields and free ghosts.
It turns out that the zero and two point functions of the $b-c$ systems
on
$\Sigma_g$ can be evaluated in terms of the operator product expansions
between these primary fields.}
\Date{October 1993}

\newsec { INTRODUCTION}
\vskip 1cm
In recent years there have been many physical applications of the theory
of affine algebraic curves, not only in the traditional field of the
string theories
\ref\strings{
D. J. Gross, P. F. Mende, {\it Phys. Lett.} {\bf 179B} (1987), 129;
{\it Nucl. Phys.} {\bf B303} (1988), 407;
R. Iengo, {\it Nucl. Phys.} {\bf B15} (Proc. Suppl.) (1990), 67;
E. Gava, R. Iengo, G. Sotkov, {\it Phys. Lett.}  {\bf 207B} (1988),
283;
D. Lebedev, A. Morozov, {\it Nucl. Phys.} {\bf B302} (1988), 63;
A. A. Belavin, V. G. Knizhnik, A. Yu. Morozov, A. M. Perelomov, {\it
JETP
Lett.} {\bf 43} (1986), 411.}, \ref\ffstr{F. Ferrari,
{\it Int. Jour. Mod. Phys.} {\bf A5} (1990),
2799.} \ref\ffstrtwo{ F. Ferrari, {\it Int. Jour. Mod. Phys.}
{\bf A7} (1992), 5131.}, \ref\jantwo{J. Sobczyk, W. Urbanik,
{\it Lett. Math. Phys.} {\bf 21} (1991), 1.},
but also in general conformal field theories
\ref\zam{Al. B. Zamolodchikov, {\it Nucl. Phys.} {\bf B285 [FS19]}
(1987), 481.}
and
very different topics like solutions of the Einstein equations
\ref\matv{ D. A. Korotkin, V. V. Matveev
{\it Leningrad Math. Jour.} {\bf 1} (1990), 379.}
integrable models
\ref\integ{I. M. Krichever, {\it Russian Math. Surveys} {\bf 32}
(6) (1977), 185; {\it Funct. Anal. Appl.} {\bf 11} (1) (1977), 12;
H. P. McKean, E. Trubowitz, {\it Comm. Pure Appl. Math.} {\bf XXIX}
(1976), 143.} and
the theory of defects \ref\zve{E. I. Zverovich, {\it Russ. Math.
Surv.} {\bf 21} (1971), 99.}.
The best known and simplest
algebraic curves are the hyperelliptic ones, so that it is not a
surprise that
a big part of the publications quoted above deals with them. However,
there
are interesting applications in which more general curves are involved,
e.g.
the conformal field theories with $Z_n$ symmetry
(\ref\knirev{V. G. Knizhnik, {\it Sov. Phys. Usp.} {\bf 32}(11)
(1989) 945.}, \ref\brzn{
M. A. Bershadsky and A. O. Radul, {\it Int. Jour. Math.
Phys.} {\bf A2} (1987) 165.})
which we investigate in the present paper.

In what follows we consider a conformal
field theory on a closed and orientable
Riemann surface $\Sigma_g$ of genus $g$. It is well known that such
$\Sigma_g$ can be
represented as an algebraic curve in ${\bf CP}_2$. Points of $\Sigma_g$
are labeled by values
of a pair of meromorphic functions $z$ and $y$ from $\Sigma_g$ onto ${\bf
CP}_1$. They are
defined up to a birational transformation. Hence, one can involve a
multivalued function
\eqn\mapping{\tilde y\equiv y\circ z:{\bf CP}_1 \rightarrow {\bf
CP}_1.}
Thus we arrived at a description (a model) of $\Sigma_g$ as a branch
covering of
the Riemann sphere. The same letter $z$ will denote a function on the
Riemann
surface in question and a coordinate on the complex plane. It is very
convenient and hopefully will not lead to any confusion. The mapping
$\tilde y(z)$
takes n values at a point except from a set of branch points $a_1,\ldots
,a_M$ where
the number of values is less then n. The local monodromy group $G$
describes the
way in which the branches of $\tilde y$ are exchanged while moving
around the
branch points. In the rest of this paper we assume (to avoid extra
purely
technical complications) that the point at infinity is not a branch
point.
Therefore, it is sufficient to study the restriction of $\tilde y$ to
the complex
plane ${\bf C}$. By means of the inverse mapping $\tilde y^{-1}$ one
can project now the
correlation functions of our
field theory on the complex plane. In order
to
define a conformal field theory, one should introduce a set of primary
fields
with rational or irrational conformal weights. One
example is
provided by the screening charges of the minimal models
\ref\minmod{
B.L. Feigin, D.B. Fuchs, {\it Funct. Anal.
Applic.}
{\bf 17}
(1983), 241; A. A. Belavin, A. M. Polyakov, A. B. Zamolodchikov,
{\it Nucl. Phys.} {\bf B241} (1984), 333; V. S. Dotsenko, V. A. Fateev,
{\it Nucl. Phys.} {\bf B240} [FS12] (1984), 312;
G. Felder, {\it Nucl. Phys.} {\bf B317} (1989), 215.}.
Unfortunately, the presence of these primary fields
modifies the monodromy behavior
of the correlation
functions \ref\bonora{
G. Felder, J. Fr\"ohlich, G. Keller, {\it Comm. Math. Phys.} {\bf 124} (1989),
647; L. Bonora, M. Matone, F. Toppan and K. Wu, {\it Phys. Lett.}
{\bf 224B} (1989) 115; Nucl. Phys. {\bf B334} (1990) 717.}
in a way which is difficult to handle on an algebraic curve.
As a consequence, the best strategy is to start from the fermionic
$b-c$ systems with integer
conformal weight $\lambda$ defined on $\Sigma_g$.
Due to their dependence on $y$, the related correlation functions become
in fact multivalued tensors on the complex plane and, using the
techniques of the Riemann monodromy problem, \knirev,
\ref\sjm{
M. Sato, T. Miwa and M. Jimbo. Holonomic quantum fields
(Kyoto U.P. Kyoto), part I; 14 (1978) p. 223; II: 15 (1979) p. 201;
III: 15 (1979) p. 577; IV: 15 (1979) p. 871; V; 16 (1980) p.
531.}, \ref\rmp{D. V. Chudnowski in: C. Bardos, D. Bessis (eds.),
Bifurcation Phenomena in Mathematical
Physics and Related Topics, D. Reidel Publishing Company 1980.},
can be expressed as
a linear combination of a finite number of multivalued tensors.
Motivated
by the analogy of these tensors with the conformal blocks of conformal
field theory, we will call them multivalued blocks.
The coefficients appearing in the linear combination mentioned above are
singlevalued functions on {\bf C}.
Moreover, the multivalued blocks should form a basis in which all the
possible multivalued behaviors compatible with the local monodromy group
$G$ of $\Sigma_g$ are represented.
Normally, the elements of this basis are chosen in such a way that they
build a set of independent solutions of the Riemann monodromy problem
\sjm, \rmp\ associated with the group $G$.
However, this is not a necessary requirement and one is allowed to
consider other bases.
Once a basis of multivalued blocks is known, it was conjectured in ref.
\ref\cmp{F. Ferrari, {\it Comm. Math. Phys.} {\bf 156} (1993) 179}
that it is possible to treat the $b-c$ systems
on an algebraic curve as a multivalued field theory on the complex
plane. The proof of this statement in the case of the $Z_n$
symmetric curves is contained in Section 2.
This is the main result of our paper. The method used for the proof
is to construct an operator formalism
\foot{We notice at this point that operator formalisms on Riemann
surfaces have been already intensively studied in the past
(see for example
\ref\opform{ C. Vafa, {\it Phys.
Lett.} {\bf 190B} (1987), 47; L. Alvarez Gaum\'e, C. Gomez, C. Reina,
{\it Phys. Lett.} {\bf 190B} (1987), 55.}) from a
point of view which is very different from the one developed here.}
for the multivalued fields in such a
way that their amplitudes coincide with the amplitudes of the $b-c$
systems on the Riemann surface. This approach
does not exploit the peculiar properties of the $Z_n$ symmetric
curves and can hopefully be extended to more general curves \ref\prep{F.
Ferrari, J. Sobczyk, W. Urbanik, in preparation.}.
\smallskip
Moreover, using the operator formalism presented here and
the fact that the solutions of the Riemann-Hilbert problem
can be expressed
\foot{
The fact that the solutions of the Riemann monodromy problem can be
described in terms of twist fields and free fermions has been first
discussed in refs. \sjm. The explicit form of the twist fields on a
$Z_n$ symmetric curve has been given in refs. \knirev\ and \brzn. The
generalization to the algebraic curves with nonabelian monodromy group
of symmetry $D_n$ has been discussed in \cmp,
\ref\pl{F. Ferrari, {\it Phys. Lett.} {\bf 277B} (1992), 423.} and
\ref\bgs{F. Ferrari, {\it Multivalued
Fields on the Complex Plane and Braid Group Statistics}, Preprint
LMU-TPW 93-24, to appear in {Int. Jour. Mod. Phys.} {\bf A}.}.
Finally, the existence of primary fields on more general algebraic
curves has been confirmed in \ffstr\
and \ref\janone{J. Sobczyk, {\it Mod. Phys. Lett.} {\bf A6}
(1991), 1103.}.}
as correlators of a conformal field theory
containing free fermions and twist fields \knirev, \sjm, \rmp,
it is possible to show a deep connection between the $b-c$
systems on an algebraic curve and the conformal field
theories on the complex plane.
Of course, these cannot be standard conformal field theories
since the amplitudes of the $b-c$ systems on $\Sigma_g$
are multivalued on {\bf C}.
This fact becomes evident in ref. \cmp, where
algebraic curves with nonabelian group of internal
authomorphisms $D_n$ are considered. In this case, in fact,
the twist fields become
nonlocal objects with a
complicated statistics \bgs.
Despite of many efforts, curves with  nonabelian
group of symmetry remain very difficult to handle, but in
the simpler situation of
a $Z_n$ group
it is indeed possible to construct a conformal field theory
on {\bf C} whose amplitudes coincide with the amplitudes of the
$b-c$ systems on $\Sigma_g$. This is done in Section 3,
where the determinant of the zero modes and the two point
function of the $b-c$ systems on a Riemann surface
are exactly reproduced from the operator product expansions
of free $b-c$ systems and twist fields on the complex plane.
\smallskip
The rest of the material presented in this paper is divided as follows.
In the conclusion some possible extensions of our operator formalism to
more general algebraic curves are discussed.
Finally, in the appendix we prove proposition two of Section 2, which
states that it is possible to compute the correlator $\langle
b(z_1)\ldots b(z_{N_b})\rangle$ of the $b-c$ systems on $\Sigma_g$
($N_b=(2\lambda-1)(g-1)$) in terms of an analogous correlator containing
only multivalued fields on the complex plane.

\vskip 1cm
\newsec{THE OPERATOR FORMALISM}
\vskip 1cm
Let us consider the $Z_n$ symmetric algebraic curves of the kind:
\eqn\zncurve{y^n=\prod_{i=1}^{nm}(z-a_i)} where
$n$ and $m$ are integers, while $z\in {\rm\bf CP}_1$.
The points $a_i\in{\rm\bf C}$ are the branch points of the curve.
$y(z)$ can be viewed both as a function on the curve $\Sigma_g$ given by
\zncurve\ and as a
multivalued function on ${\rm\bf CP}_1$ with $n$ branches (strictly
speaking it should be then denoted as $\tilde y$ as in \mapping\ but in
what follows we shall omit writing $\tilde {}$).
The genus of the surface $\Sigma_g$ is given by:
\eqn\genus{g=1-n+{nm(n-1)\over 2}.}
On $\Sigma_g$ we consider the theory of fermionic free  $b-c$ fields of
spin $\lambda\in{\rm\bf Z}$:
\eqn\action{S_{\rm bc}=\int_{\Sigma_g}d^2\xi\left(b\bar\partial c+{\rm
c.c.}\right)}
where $\xi$ and $\bar \xi$ are coordinates on $\Sigma_g$.
Such theories on algebraic curves have already been discussed in refs.
\ffstr, \pl, \cmp, \janone, \jantwo.
(see also refs. \bonora\ and \ref\others{
L. Dixon, D. Friedan, E. Martinec, S. Shenker,
{\it Nucl. Phys.} {\bf B282} (1987), 13;
S. Hamidi, C. Vafa, {\it Nucl. Phys.} {\bf B279} (1987), 465;
J. J. Atick, A. Sen, {\it Nucl. Phys.} {\bf B286} (1987), 189;
E. Guadagnini, M. Martellini, M. Mintchev, {\it Jour. Math. Phys.}
{\bf 31} (1990), 1226;
J. Sobczyk, {\it Current-Current Correlation Functions on Algebraic
Curves}, Preprint ITP UWr-791/91; {\it Insertion at Infinity in
Conformal Field Theories on Algebraic Curves}, Preprint ITP UWr 792/91.}
where multivalued field theories have been
treated in different contexts).
\smallskip
The theory defined by \action\ can also be viewed as a multivalued field
theory on the complex plane. A problem arises if it is possible to
define both theories in such a way that correlation functions produced
by them are related just by the projection from $\Sigma_g$ to ${\rm\bf
CP_1}$. In the second theory $y(z)$ is a multivalued function with $n$
branches denoted by $y^{(l)}(z)$, $l=0,\ldots,n-1$. Any tensor
$T$ on $\Sigma_g$ can be expressed by means of $z$, $y(z)$ and $dz$
and when projected becomes multivalued as well:
$T^{(l)}(z)\equiv T(z,y^{(l)}(z))$.
In the following we will usually omit the branch index $l$ for
simplicity.
To establish a working operator formalism on $\Sigma_g$, we have
to find a suitable basis in which the $b-c$ fields should be
expanded. This is provided by techniques used in the Riemann
monodromy problem (see ref.
\ref\by{B. Blok and S. Yankielowicz, {\it Nucl. Phys.}
{\bf B321} (1989) 327.
B. Blok and S. Yankielowicz, {\it Phys. Lett.}
{\bf 226B} (1989) 279.}
for more details).
First of all, we need a set of functions $F_k(z,y(z))$ on $\Sigma_g$,
$k=0,\ldots,n-1$
being independent in the following way:
They fulfill the condition $F_k(z,y(z))/F_{k'}(z,y(z))=f(z)$ only if
$k=k'$, where $f(z)$ is a single-valued function on ${\rm\bf CP}_1$.
When $k=k'$, of course, $f(z)=1$.
It is possible to show that the basis of functions $F_k(z,y(z))$ can
have
only $n$ elements.
Any meromorphic function $g(z,y(z))$ on $\Sigma_g$ can be
now represented as a linear combination of the functions $F_k(z,y(z))$
with coefficients being single-valued functions on ${\rm\bf CP}_1$.
There is still a large freedom in choosing particular bases $F_k$. An
extra requirement we impose is that for a given value of $\lambda$ we
would like to find two bases: $f_k$ and $\phi_k$ such that
\eqn\kzw{K(z,w)dz^\lambda dw^{1-\lambda}={1\over
z-w}\sum\limits_{k=0}^{n-1} f_k(z)\phi_k(w)dz^\lambda dw^{1-\lambda}}
where the object on the LHS of \kzw\ has just (as a differential in $z$)
one single pole at $w$ with residue n and presents itself as a basic
building block of correlation function on Riemann surfaces (for details
see \ffstr ).

The solution of the classical equations of motion for \action\
\eqn\classeq{\bar\partial b=\bar\partial c=0}
are meromorphic tensors, which can be expanded in the following basis:
\eqn\bdz{b(z)dz^\lambda=\sum\limits_{k=0}^{n-1}\sum\limits_{i=-\infty}
^\infty b_{k,i}z^{-i-\lambda}f_k(z)dz^\lambda}
\eqn\cdz{c(z)dz^{1-\lambda}=
\sum\limits_{k=0}^{n-1}\sum\limits_{i=-\infty}^\infty
 c_{k,i}z^{-i+\lambda-1}\phi_k(z)dz^{1-\lambda}}
where
\eqn\fkn{f_k(z)dz^\lambda ={dz^\lambda\over[y(z)]^{-k+\lambda(n-1)}}
\qquad\qquad\qquad k=0,\ldots,n-1}
\eqn\phikn{\phi_k(z)dz^{1-\lambda}=
{dz^{1-\lambda}\over[y(z)]^{k-\lambda(n-1)}}\qquad
\qquad\qquad k=0,\ldots,n-1.}
The expansions \bdz\ and \cdz\ resemble the local expansions of the
$b-c$ fields on the sphere. The only difference is provided by the
presence of the functions $f_k(z)$ and $\phi_k(z)$. This is the
contribution coming from the topology of the Riemann surface. It is easy
to check that $f_k(z)$ and $\phi_k(z)$
form two different basis of rationally
independent functions on $\Sigma_g$ satisfying the requirement \kzw .
Eqs. \bdz\ and \cdz\ represent the most general possible expansions of
the fields. Now we quantize the $b-c$ systems using these bases.
In order to do it we have to postulate some anticommutation relations
for classical degrees of freedom $b_{k,i}$ and $c_{k,i}$.
It is convenient to split the fields in their components
$b_k(z)$ and $c_k(z)$:
\eqn\splitting{
b(z)=\sum\limits_{k=0}^{n-1}b_k(z)\qquad\qquad\qquad
c(z)=\sum\limits_{k=0}^{n-1} c_k(z)}
where
\eqn\bkdz{b_k(z)dz^\lambda=f_k(z)
\sum\limits_{i=-\infty}^{\infty}b_{k,i}
z^{-i-\lambda}dz^\lambda}
\eqn\ckdz{c_k(z)dz^{1-\lambda}=\phi_k(z)
\sum\limits_{i=-\infty}^\infty
c_{k,i}z^{-i+\lambda-1}dz^{1-\lambda}}
\eqn\commrel{\{b_{k,i},c_{k',i'}\}=\delta_{kk'}\delta_{i+i',0}.}
Moreover, we introduce vacua $|0\rangle_k$, on which the operators
$b_{k,n}$ and $c_{k',m}$ act.
The total vacuum of the $b-c$ systems on $\Sigma_g$ is given by
\eqn\totalvacuum{|0\rangle=\otimes_{k=0}^{n-1}|0\rangle_k.}
{}From now on, we will suppose that
$\lambda>1$. The case $\lambda=1$ can be treated in an analogous
way with an obvious complication coming from the $c$ field zero mode.
We also consider only Riemann surfaces of genus $g\ge 2$ in order to
avoid the exceptional algebraic curves.
The zero modes occurring in a given $k-$component $b_k(z)$ of the fields
$b(z)$ are of the form:
\eqn\zeromodes{\Omega_{k,j}dz^\lambda={z^{j-1}dz^\lambda
\over[y(z)]^{-k+\lambda(n-1)}}\qquad\qquad\qquad j=1,\ldots,N_{b_k}}
where $N_{b_k}=-2\lambda+1+\lambda(n-1)m-km$.
They correspond to the operators $b_{k,i}$ of eq. \bkdz\ with
$\lambda-\lambda(n-1)m +km\le i\le -\lambda$.
It is easy to check that
$\sum\limits_{k=0}^{n-1}N_{b_k}=(2\lambda-1)(g-1)$, giving exactly the
Riemann-Roch theorem for a Riemann surface of genus $g$, where $g$ is
given by eq. \genus.    A treatment of the $b-c$ systems on $\Sigma_g$
can be now performed using the techniques exploited in the case of
genus zero.
The annihilation operators are components of the fields $b(z)$ and
$c(z)$ with negative powers of $z$. Since there is the splitting in
$n$-component of the fields showed by eqs. \bkdz-\commrel, this implies
that:
\eqn\ban{b^-_{k,i}|0\rangle_k\equiv b_{k,i}|0\rangle_k=0
\qquad\qquad\qquad\left\{\eqalign{k=&0,\ldots,n-1\cr i
\ge& 1-\lambda\cr}\right.}
\eqn\can{c^-_{k,i}|0\rangle_k\equiv c_{k,i}|0\rangle_k=0
\qquad\qquad\qquad\left\{\eqalign{k=&0,\ldots,n-1\cr i\ge&
\lambda\cr}\right.}
We also introduce the "out" vacua ${}_k\langle 0|$ requiring:
\eqn\bcrea{{}_k\langle0| b^+_{k,i}\equiv {}_k\langle 0|b_{k,i}=0
\qquad\qquad\qquad\left\{\eqalign{k=&0,\ldots,n-1\cr i
\le& -\lambda-N_{b_k}\cr}\right.}
\eqn\ccrea{{}_k\langle 0|c^+_{k,i}\equiv {}_k\langle 0|c_{k,i}=0
\qquad\qquad\qquad\left\{\eqalign{k=&0,\ldots,n-1\cr i\le&
\lambda-1\cr}\right.}
By definition, the normal ordering of the product of two fields
$b(z)c(w)$
consists in putting the annihilation operators $b^-_{k,n}$ and
$c^-_{k,n}$
to the right of the creation operators $b^+_{k,n}$ and $c^+_{k,n}$.
Applying the definition \bkdz\ and \ckdz\ of the fields $b$ and $c$ and
exploiting the commutation relations \commrel, it is possible to see
that (we assume that $|w/z|<1$):
\eqn\bkzckwnorm{b_k(z)c_k(w)=:b_k(z)c_k(w):+{1\over z-w}f_k(z)\phi_k(w)}
\eqn\ckwbkznorm{c_k(z)b_k(w)=:c_k(z)b_k(w):+{1\over z-w}f_k(w)\phi_k(z)
.}
Moreover, in order to take into account also the zero modes, we have to
impose the following conditions on the vacua $|0\rangle_k$:
\eqn\vaccond{{}_k\langle 0|0\rangle_k=0\qquad{\rm if}\ N_{b_k}\ne 0;
\qquad\qquad
{}_k\langle 0|\prod\limits_{i=1}^{N_{b_k}}b_{k,i}|0\rangle_k=1.}
With the above definitions, the proof of the proposition below becomes
straightforward and we do not report it.\smallskip\noindent
{\bf Proposition 1}:
\eqn\propone{{}_k\langle 0|b_k(z_1)\ldots b_k(z_{N_{b_k}})|0\rangle_k=
{\rm Det}\left|
\Omega_{k,j}(z_i)\right|\qquad\qquad\qquad i,j=1\ldots,N_{b_k}}
where $||\Omega_{k,j}(z_i)||$ is a matrix of the zero modes given by eq.
\zeromodes.\smallskip
The second step in order to set up an operator formalism is the
following:\smallskip\noindent
{\bf Proposition 2}:
\eqn\proptwo{\langle 0|b(z_1)\ldots b(z_{N_b})|0\rangle={\rm
det}\left|\Omega_I(z_J)\right|}
where the vacuum $|0\rangle$ and the field $b(z)$
have been already defined in
eqs. \totalvacuum\ and \bdz, \cdz\ respectively.
Moreover $I,J=1,\ldots,\sum\limits_{k=0}^{n-1}N_{b_k}= N_b$, $N_b$
describing the total number of the zero modes.
Finally the $\Omega_I(z)dz^\lambda$ represent all the possible zero
modes with spin $\lambda$:
$$\Omega_I(z)dz^\lambda\in\left\{\Omega_{k,i}(z)dz^\lambda|1\le i\le
N_{b_k}, 0\le k\le n-1\right\}.$$
Proposition 2 is proved in the Appendix A.\smallskip
Now we are ready to compute the propagator of the $b-c$ systems using
the bases \bdz\ and \cdz.
In our operator formalism, the usual propagator of the $b-c$ systems is
the following ratio of correlators:
\eqn\propoper{G_\lambda(z,w)={\langle 0|b(z)c(w)\prod\limits_{I=1}^{N_b}
b(z_I)
|0\rangle\over
\langle 0|\prod\limits_{I=1}^{N_b}b(z_I)|0\rangle}}
where the vacuum $|0\rangle$ and the fields $b$ and $c$ are provided by
eqs.
\totalvacuum\ and \splitting\ respectively.
The denominator of eq. \propoper\ can be easily evaluated using eq.
\proptwo .
To compute the numerator, instead, we have to use the normal ordering
given in eqs. \bkzckwnorm\ and \ckwbkznorm . The normal ordering of two
fields
$b$ and $c$ becomes:
\eqn\normord{b(z)c(w)=:b(z)c(w):+K(z,w)dz^{\lambda}dw^{1-\lambda}.}
Since $K(z,w)$ is a multivalued tensor,
one should specify in which branches in $z$ and $w$ the
singularity arises. It turns out that the pole at $z=w$ occur only if
the
branches of $K(z,w)$ in $z$ and $w$ are the same (see for example
\ffstr).
The fact that $K(z,w)dz^\lambda dw^{1-\lambda}$ is a tensor with only
one singularity in $z=w$ will play an important role in the following.
The reason is that a tensor of this kind is one of the building blocks
in the construction of the $n-$point functions of the $b-c$ systems on
an algebraic curve. The other building blocks are the zero
modes.\smallskip
Now we are ready to evaluate the propagator given in eq. \propoper .
We suppose that the fields are radial ordered, i.e.
$$|z|>|w|>|z_1|\ldots>|z_{N_b}|.$$
After simple calculations we find:
$${\langle 0|b^{(l)}(z)c^{(l')}(w)\prod\limits_{I=1}^{N_b}b^{(l_I)}(z_I)
|0\rangle\over
\langle 0|\prod\limits_{I=1}^{N_b}b^{(l_I)}(z_I)|0\rangle}=
K(z,w)^{(ll')}dz^\lambda
dw^{1-\lambda}+$$
\eqn\propint{\sum\limits_{J=1}^{N_b}(-1)^JK_\lambda^{(l_Jl')}(z_J,w)
{\langle b^{(l_1)}(z_1)\ldots
b^{(l_{J-1})}(z_{J-1})b^{(l)}(z)b^{(l_{J+1})}
(z_{J+1})
\ldots b^{(l_{N_b})}(z_{N_b})\rangle\over
\langle b^{(l_1)}(z_1)\ldots b^{(l_{N_b})}(z_{N_b})\rangle}}
where $l$, $l'$ and $l_I,l_J$ denote the branches of fields and tensors
in the variables $z$, $w$, $z_I, z_J$ respectively.
The residual correlation function in eq. \propint\ contain products of
$N_b$ fields $b$ and therefore can be easily computed by means of eq.
\proptwo.
The final result is the following propagator:
$${\langle 0|b^{(l)}(z)c^{(l')}(w)\prod\limits_{I=1}^{N_b}b^{(l_I)}(z_I)
|0\rangle\over
\langle 0|\prod\limits_{I=1}^{N_b}b^{(l_I)}(z_I)|0\rangle}=$$
\eqn\propfin{{
{\rm det}\left|\matrix{\Omega_1^{(l)}(z)&\ldots&
\Omega_{N_b}^{(l)}(z)&K_\lambda^{(ll')}(z,w)\cr
\Omega_1^{(l_1)}(z_1)&\ldots&
\Omega_{N_b}^{(l_1)}(z_1)&K_\lambda^{(l_1l')}(z_1,w)\cr
\vdots&\ddots&\vdots&\vdots\cr
\Omega_1^{(l_{N_b})}(z_{N_b})&\ldots&
\Omega_{N_b}^{(l_{N_b})}(z_{N_b})&K_\lambda^{(l_{N_b}l')}(z_{N_b},w)
\cr}\right |\over
{\rm det}\left|\Omega_I(z_J)\right|}.}
In the above formula we have introduced the indices denoting the
branches in order to facilitate the comparison
with the propagator of the $b-c$ systems on the $Z_n$
symmetric curves that has been found in \ref\ferbos{F. Ferrari, {\it
Jour. Math. Phys.} {\bf 32} (1991), 2186.}
using the method of the fermionic construction.
It is easy to see that the two results coincide showing that the
operator formalism here established is able to reproduce the two
point function on a $Z_n$ symmetric algebraic curve.
Moreover, starting from eq. \propfin, we can compute all the other
$n-$point functions applying the Wick theorem. The Wick theorem for the
$b-c$ systems has been rigorously studied in
\ref\raina{
A. K. Raina, {\it Helv. Phys. Acta} {\bf 63} (1990), 694.}
and it
is valid also in our case.
One can check it inductively starting from eq. \propfin\ and supposing
that the Wick theorem has been checked for the correlator
$$G_{N-1,M-1}(z_1,\ldots,z_{N-1};w_1,\ldots,w_{M-1})=\langle
0|b(z_1)\ldots b(z_{N-1})c(w_1)\ldots c(w_{M-1})|0\rangle$$
with $N-M=N_b$. Then, using eq. \normord\ we obtain:
$$\langle 0|b(z_N)c(w_M)
b(z_1)\ldots b(z_{N-1})c(w_1)\ldots
c(w_{M-1})|0\rangle=$$
\eqn\brak{\sum\limits_{i=1}^M(-1)^i K_\lambda(z_N,w_i)
G_{N-1,M-1}(z_1,\ldots,z_{N-1};w_1,\ldots,w_{i-1},w_{i+1},\ldots w_{M-1}
).}
All the other possible contractions vanish due to the fact that the Wick
theorem holds by hypothesis in the case of any product containing
$N-1$ fields $b$ and $M-1$ fields $c$.
As an upshot we obtain:
$$
<\prod\limits_{s=1}^Mb^{(l_s)}(z_\rho)\prod\limits_{t=1}^Nc^{(l'_t)}
(w_t)>=$$
\eqn\bclambda{{\rm det}\left|\matrix{\Omega_1^{(l_1)}(z_1)&\ldots&
\Omega_{N_b}^{(l_1)}(z_1)&K_\lambda^{(l_1l'_1)}(z_1,w_1)&
\ldots&K_\lambda^{(l_1l_N')}(z_1,w_N)\cr
\vdots&\ddots&\vdots&\vdots&\ddots&\vdots\cr
\Omega_1^{(l_M)}(z_M)&\ldots&
\Omega_{N_b}^{(l_M)}(z_M)&K_\lambda^{(l_Ml'_1)}(z_M,w_1)&
\ldots&K_\lambda^{(l_Ml_N')}(z_M,w_N)\cr}\right |}
where $M-N=(2\lambda-1)(g-1)=N_b$.
The tensor $K_\lambda^{(ll')}(z,w)$ has spurious poles in the limit
$w\rightarrow\infty$. However one can show as in
\ffstr\
and
\ref\bi{
M. Bonini and R, Iengo, {\it Int. Jour. Mod. Phys.} {\bf A3}
(1988) 841.}
that these poles do not contribute to the determinant \bclambda.
The important fact to be noted here is that both eqs. \propfin\ and
\bclambda\ were evaluated using the operator formalism explained above,
which can be considered for this reason
as an operator formalism  valid on the Riemann surfaces
of any genus with $Z_n$ symmetry.

\vskip 1cm
\newsec{THE $b-c$ SYSTEMS ON A RIEMANN SURFACE AS A CONFORMAL FIELD
THEORY}
\vskip 1cm
Using the results found in the previous section, we express now the
correlation functions of free fermionic $b-c$ systems on a Riemann
surface by means of a suitable conformal field theory on the sphere.
The content of primary fields is the following.
First of all, we need a set of $n$ free $b-c$ fields:
\eqn\btk{\tilde b_k(z)dz^\lambda=\sum\limits_{i=-\infty}^\infty
\tilde b_{k,i}z^{-i-\lambda} dz^{\lambda}}
\eqn\ctk{\tilde
c_k(z)dz^{1-\lambda}=\sum\limits_{i=-\infty}^{\infty}\tilde c_{k,i}z^{-i+
\lambda -1} dz^{1-\lambda}}
with
\eqn\ordertilde{\tilde b_k(z)\tilde c_{k'}(w)\sim{\delta_{kk'}
\over {z-w}}+:\tilde b_k(z)\tilde c_{k'}(w):.}
Moreover, we introduce the spin fields $V_k(a_l)$, $l=1,\ldots,mn$ such
that:
\eqn\bkvk{\tilde b_k(z)V_k(a_l)\sim (z-a_l)^{-q_{k,a_l}}:\tilde
b_k(z)V_k(a_l): +\ldots}
\eqn\ckvk{\tilde c_k(z)V_k(a_l)\sim (z-a_l)^{q_{k,a_l}}:\tilde
c_k(z)V_k(a_l): +\ldots}
where:
\eqn\qkal{q_{k,a_l}={\lambda(n-1)-k\over n}\qquad\qquad\qquad
k=0,\ldots,n-1.}
Theories with
spin fields of that kind have been already discussed in refs. \knirev\
and \brzn.
In the first part of this section, therefore, we will follow \brzn\ in
order to compute the correlator:
\eqn\corrsk{s_k={}_k\langle\tilde 0|\prod\limits_{i=1}^{N_{b_k}}\tilde
b_k(z_i)\prod\limits_{l=1}^{mn} V_k(a_l)|\tilde 0\rangle_k.} The result
will be
an equation which is the equivalent of eq. \propone.
We notice that $s_k$ has no singularities in the variables $z_i$,
therefore it should be proportional to a product of the zero modes
\zeromodes.
The correlator $s_k$ can be computed using the energy momentum tensor
method, i.e. exploiting the fact that if $\phi_1(z_1)\ldots\phi_N(z_N)$
are primary fields then the following equation holds:
\eqn\emope{\langle
T(z)\phi_1(z_1)\ldots\phi(z_N)\rangle =\sum\limits_{i=1}^N
\left({\Delta_i\over(z-z_i)^2} +{1\over z-z_i}\partial_{z_i}\right)
\langle \phi_1(z_1)\ldots\phi(z_N)\rangle}
where $T(z)$ is the energy momentum tensor.
In our case it is splitted in
$n$ components $T_k(z)$, $k=0,\ldots,n-1$, where:
\eqn\emten{T_{k,zz}(z)=-\lambda\partial_z\tilde b_k\tilde c_k
+(1-\lambda)\tilde b_k\partial_z\tilde c_k.}
Therefore, instead of eq. \emope, we have to evaluate:
\eqn\tkexp{\langle\langle T_k(z)\rangle\rangle\equiv{
{}_k\langle\tilde
0|T_k(z)\prod\limits_{i=1}^{N_{b_k}} \tilde
b_k(z_i)\prod\limits_{l=1}^{nm}V_k(a_l)|\tilde 0\rangle_k\over
{}_k\langle\tilde
0|\prod\limits_{i=1}^{N_{b_k}} \tilde
b_k(z_i)\prod\limits_{l=1}^{nm}V_k(a_l)|\tilde 0\rangle_k}.}
As a first step, we derive the following Green function:
\eqn\confblocks{F_k(z,z_0)={{}_k\langle\tilde
0|\tilde b_k(z)\tilde c_k(z_0)\prod\limits_{i=1}^{N_{b_k}} \tilde
b_k(z_i)\prod\limits_{l=1}^{nm}V_k(a_l)|\tilde 0\rangle_k
\over
{}_k\langle\tilde
0|\prod\limits_{i=1}^{N_{b_k}} \tilde
b_k(z_i)\prod\limits_{l=1}^{nm}V_k(a_l)|\tilde 0\rangle_k}.}
As shown in refs. \pl\ and \cmp\ in the more general case of algebraic
curves with nonabelian group of symmetry $D_n$, the total ghost charge
in the
numerator and in the denominator of eq. \confblocks\ is zero, so that
$F_k(z,z_0)$ is well defined.
In particular, let us notice that the numerator of eq. \confblocks\ is
the
only possible correlator with the insertion of only one $c$ field
that does not vanish. This fact will be useful below when we will
compute the propagator of the $b-c$ systems on $\Sigma_g$ in terms of
the Green function $F_k(z,z_0)$.
As noticed in \brzn\ the Green functions \confblocks, which are tensors
in $z$ and $z_0$, are completely determined by their poles and zeros in
$z$, $z_0$ and $z_i$. The upshot is (see \brzn\ and \knirev\ for more
details):
\eqn\fkzzz{F_k(z,z_0)={f_k(z)\phi_k(z_0)\over
z-z_0}\prod\limits_{i=1}^{N_{b_k}}\left({z-z_i\over z_0-z_i}\right).}
We note that $F_k(z,z_0)$ is normalized in such a way that $\lim
\limits_{z\to
z_0}{F_k(z,z_0)}=1$. The Green functions $F_k(z,z_0)$ will play the role
of the multivalued blocks in our conformal field theory induced by the
$b-c$ systems on $\Sigma_g$.
In fact, $F_k(z,z_0)$ is proportional to the function $f_k(z)$ given
in eq. \fkzzz\ which describes all the possible monodromy properties
compatible with the local monodromy group $Z_n$ of $\Sigma_g$.
Now it is possible to evaluate $\langle\langle T_k(z)\rangle\rangle$
using the explicit form of the $F_k(z,z_0)$.
As a matter of fact, from eq. \emten\ we get:
\eqn\tkzbb{\langle\langle T_k(z)\rangle\rangle=\lim_{z\to
z_0}\left(-\lambda \partial_z
F_k(z,z_0)+(1-\lambda)\partial_{z_0}F_k(z,z_0) -{1\over (z-z_0)}\right)
.}
Moreover, eq. \emope\ says that the residues of $\langle\langle
T_k(z)\rangle\rangle$  in $z=a_l$ are:
\eqn\residue{{\rm Res}\langle\langle
T_k(z)\rangle\rangle_{z=a_l}=\partial_{a_l}{\rm log}\left({}_k\langle
\tilde 0|\prod\limits_{i=1}^{N_{b_k}} \tilde
b(z_i)\prod_{l=1}^{nm}V_k(a_l)|\tilde 0\rangle_k \right).}
The left hand side becomes explicitly known after having combined with
eqs \fkzzz\ and \tkzbb.
Finally, eq. \residue\ can be solved as in \brzn\ giving:
\eqn\propthree{s_k={\rm
det}|\Omega_{k,j_k}(z_{i_k})|\prod\limits_{l=1}^{nm}\prod\limits_{l'\ne
l=1}^{nm} (a_l-a_{l'})^{q_{k,a_l}q_{k,a_{l'}}}.}
This is the promised equivalent of eq. \propone.\smallskip
We would like to show that the above conformal field theory
with additional $Z_n$ symmetry is equivalent to the
$b-c$ systems on the Riemann surface $\Sigma_g$ defined in eq. \zncurve.
To this purpose, we introduce the total vacuum
\eqn\tvac{|\tilde 0\rangle=\prod\limits_{k=0}^{n-1}|\tilde 0\rangle_k}
and the total fields:
\eqn\fieldtot{\tilde b(z)=\sum\limits_{k=0}^{n-1}\tilde
b_k(z)\qquad\qquad\qquad \tilde c(z)=\sum\limits_{k=0}^{n-1}\tilde
c_k(z).}
Finally, we define the spin fields
\eqn\spinfdstot{V(a_l)=\prod\limits_{k=0}^{n-1}V_k(a_l).}
As a first step, we prove the following equation:
\eqn\propfour{\langle\tilde 0|\tilde b(z_1)\ldots\tilde
b(z_{N_b})\prod\limits_{l=1}^{nm}V(a_l)|\tilde 0\rangle= {\rm
det}|\Omega_I(z_J)|\prod\limits_{l=1}^{nm}\prod\limits_{l'\ne
l=1}^{nm}(a_l-a_{l'})^{\gamma_{ll'}}}
where $\gamma_{ll'}=\sum\limits_{k=0}^{n-1}q_{k,a_l}q_{k,a_{l'}}$
and $I,J=1,\ldots,N_b$.
Eq. \propfour\ can be verified in the same way as eq. \proptwo .
The only difference is that instead of eq. \propone\ we have to exploit
now eq. \propthree\ in order to compute the correlators $s_k$.
This is the reason for which eq. \proptwo\ and eq. \propfour\ differ by
the factor $\prod\limits_{l=1}^{nm}\prod\limits_{l'\ne
l=1}^{nm}(a_l-a_{l'})^{\gamma_{ll'}}$.\smallskip
Now we are ready to compute the propagator \propfin\ of the $b-c$
systems on a Riemann surface in terms of the multivalued blocks \fkzzz.
We have to evaluate a correlator of the kind
\eqn\propiniz{
\langle\tilde 0|\tilde c(w)\prod\limits_{I=1}^{N_b+1}\tilde b(z_I)
\prod\limits_{l=1}^{nm}V(a_l)|\tilde 0\rangle=
\sum\limits_{k=0}^{n-1}\langle\tilde 0|\tilde c_k(w)
\prod\limits_{I=1}^{N_b+1}\tilde
b(z_I)\prod\limits_{l=1}^{nm}V(a_l)|\tilde 0\rangle.}
Following a strategy similar to that applied in the Appendix A, we
use the expansions \fieldtot\ and then
apply the Lemma 1. We observe that for each
value of $k$ there is only one partition of the number $N_b+1$ giving a
non-zero contribution. Taking into account \tvac\ the structure of the
expression we obtain is the following. It is a product of (n-1)
correlators of the type \corrsk\ multiplied by a correlator present in
the numerator of \confblocks . These (n-1) correlators can be calculated
using the formula \propthree. The last correlator to be computed in eq.
\propiniz\ can be represented in the
following very convenient form:
$${}_k\langle\tilde 0|\tilde c_k(w)\prod\limits_{i_k=\alpha (k)}^{\beta
(k)}\tilde b_k(z_{\sigma(i_k)})\prod_{l}V_k(a_l)|\tilde 0\rangle_k=$$
\eqn\partuno{=-\prod\limits_{l=1}^{nm}\prod\limits_{l'\ne
l=1}^{nm} (a_l-a_{l'})^{q_{k,a_l}q_{k,a_{l'}}}
 \sum\limits_{i_k=\alpha (k)}^{\beta (k)}{\phi_k(w)f_k(z_{
\sigma(i_k)})\over z_{\sigma(i_k)}
-w}(-1)^{i_k}{\rm det}|\Omega_{k,j_k}(z_{\sigma(l_k)})|}
where
$$\qquad\qquad
l_k\in\{\alpha (k),\ldots,\beta (k)\}\qquad{\rm with}\qquad l_k\ne i_k.
$$
Here $\alpha(k)=1+\sum\limits_{m=1}^{k-1}N_{b_m}$ and $\beta(k)=
\alpha(k)+N_{b_k}$. The notation is taken from the Appendix A with a
minor modification due to the presence of the field $\tilde c_k$ in
\partuno . When $k=0$, the sum in $\alpha(k)$ is by definition zero.
We notice that a priori we should put a sign $(-1)^{\alpha(k)}$
in the LHS of eq. \partuno. This comes from the fact that, before
to compute the correlator \partuno, we have to commute $\alpha(k)$ times
the field $\tilde c_k(w)$ with the $b-$fields. However, this sign
cancels with an equal factor appearing in the RHS, coming from the fact
that the sum over $i_k$ starts from $\alpha(k)$.
Using eq. \partuno\ we obtain:
$$\langle 0|\tilde c(w)\prod\limits_{I=1}^{N_b+1}\tilde b(z_I)
\prod\limits_l V(a_l)|0\rangle=-\prod\limits_{l=1}^{nm}\prod\limits_
{l'\ne l=1}^{nm}(a_l-a_{l'})^{\gamma_{ll'}}\times$$
\eqn\czesc{\sum\limits_{k=0}^{n-1}
\sum\limits_{\sigma}{\rm sign}(\sigma)
\prod\limits_{p\neq k}^{n-1}
{\rm det}|\Omega_{p,j_p}(z_{\sigma(i_p)})|
\sum\limits_{i_k=\alpha (k)}^{\beta (k)}{\phi_k(w)f_k(z_{\sigma(
i_k)})\over z_{\sigma(i_k)}-w}(-1)^{i_k}{\rm det}|\Omega_{k,j_k}(z_{
\sigma(l_k)})|.}
It is now useful to observe that:
$$\sum\limits_{i_k=\alpha (k)}^{\beta (k)}{\phi_k(w)f_k(z_{\sigma (
i_k)})\over z_{\sigma (i_k)}-w}(-1)^{i_k}{\rm
det}|\Omega_{k,j_k}(z_{\sigma (l_k)})|=$$
\eqn\splitprop{=(-1)^{\alpha(k)}{\rm det}\left|\matrix{
{\phi_k(w)f_k(z_{\sigma (\alpha
(k))})
\over z_{\sigma (\alpha (k))}-w}&
\Omega_{k,j_1}(z_{\sigma (\alpha
(k))})&
\ldots&\Omega_{k,j_{N_{b_k}}}(z_{\sigma (\alpha
(k))})\cr\vdots&\ddots&\vdots&\vdots\cr
{\phi_k(w)f_k
(z_{\sigma (\beta (k))})\over z_{\sigma (\beta (k))}-w}&
\Omega_{k, j_1}(z_{\sigma (\beta (k))})&
\ldots&\Omega_{k,j_{N_{b_k}}}(z_{\sigma (\beta (k))})\cr}\right |.}
It is easy to see that the factor $(-1)^{\alpha(k)}$ comes from the fact that
the sum in the LHS of eq. \splitprop\ begins from $\alpha(k)$.
Using now the Lemma 2 from the Appendix A we obtain
$$\langle\tilde 0|\tilde c(w)\prod\limits_{I=1}^{N_b+1}\tilde b(z_I)
\prod\limits_l V(a_l)|\tilde 0\rangle=-
\prod\limits_{l=1}^{nm}\prod\limits_
{l'\ne l=1}^{nm}(a_l-a_{l'})^{\gamma_{ll'}}$$
\eqn\koniec{\sum\limits_{k=0}^{n-1}(-1)^{\alpha(k)}
{\rm det}\left|\matrix{\Omega_1(z_1)&\ldots&
{\phi_k(w)f_k(z_1)\over z_1-w}&
\Omega_
{\alpha (k)}(z_1)&\ldots&
\Omega_{N_b}(z_1)\cr
\vdots&\ddots&\vdots&\vdots&\ddots&\vdots\cr
\Omega_1(z_{N_b+1})
&\ldots&{\phi_k(w)f_k(z_{N_b+1})\over
z_{N_b+1}-w}&\Omega_{\alpha (k)}(z_{N_b+1})&
\ldots&\Omega_{N_b}(z_{N_b+1})\cr}\right|.}
It is not difficult to convince oneself that with the help of eq. \kzw\
one can reproduce the formula
\propfin\ up to the term coming from the ramification points (closely
related to the chiral determinant of the appropriate Dirac operator).
As a matter of fact, the wrong position of the column containing the
tensors ${\phi_k(w)f_k(z_I)/
(z_I-w)}$, $I=1,\ldots,N_b$, is compensated by the factor $(-1)^{\alpha(k)}$.
Moreover, the overall minus sign in the RHS of eq. \koniec\ takes into
account of the fact that, with respect to eq. \propfin, the field $c(w)$
has been once commuted with a $b-$field.
Thus we have proved that the two point function of the $b-c$ systems can
be computed starting from the conformal field theory with $Z_n$ symmetry
described at the beginning of this section.

\vskip 1cm
\newsec{CONCLUSIONS}
\vskip 1cm
In this paper we have shown that the $b-c$ systems on a
Riemann surface are equivalent to a multivalued theory of ghosts defined
on the
complex plane, in which the fields are expanded in the basis
given by eqs. \bkdz\ and \ckdz. It is clear that this basis has
zeros and poles at the projections on the Riemann surface
of the points $z=0$ and $z=\infty$ respectively.
In this sense, our basis represents an example of the
generalized Krichever-Novikov bases \ref\kn{ I.
M. Krichever, S. Novikov, {\it Funkt. Anal. i Pril.} {\bf 211} (1987),
46.} of the kind discussed in \ref\kngen{R. Dick, {\it Lett. Math.
Phys.} {\bf 18} (1989), 255; M. Schlichenmaier, {\it Lett. Math. Phys.}
{\bf 19} (1990), 151.}. In
the basis \bkdz\ and \ckdz\ the
derivation of the amplitudes of the $b-c$ systems becomes simpler, but
we believe  that it is possible to set up an operatorial
formalism like that of section 2 also starting from
any other equivalent basis provided that \kzw\ is satisfied.
Obviously, the operator formalism of section 2 is a general one
and can be extended to other curves.
If this is true, it implies
that the $b-c$ systems on an arbitrary Riemann surface
are equivalent to a multivalued field theory on the complex plane.
The next step in order to verify this conjecture is provided by
the $D_n$ symmetric curves discussed in ref.
\cmp\ and by the example of a quartic of genus three \ffstr,
\jantwo.
\smallskip
On the other side, in section 3 we have also
shown how the amplitudes of the $b-c$ systems on the $Z_n$
symmetric curves are determined by the operator product expansions
of the primary fields
of eqs. \ordertilde, \bkvk\ and \ckvk.
Explicit examples have been given in the case of the zero and
two point functions (eqs. \propfour\ and \koniec .
The extension of this result also to the $D_n$ symmetric curves of ref.
\cmp\ seems to be feasible.
This is an interesting class of algebraic curves because
the exchange algebra between the twist fields
provides an example of nonabelian statistics involving
multiparameter Yang$-$Baxter matrices \bgs.\smallskip

\vskip 1cm
\appendix {A}{}
\vskip 1cm
The proof of the proposition 2 becomes quite straithforward once two
basic lemmas are established.\medskip\noindent
{\bf Lemma 1}\smallskip
\eqn\super{\prod\limits_{i=1}^N (\sum\limits_{j=1}^Ma_{i,j})=
\sum\limits_{r_1+\ldots +r_M=N}
\sum\limits_{\sigma}{\rm sgn}(\sigma )\prod\limits_{k=1}^N\prod\limits_
{l=\alpha (k)}^{\beta (k)}a_{\sigma (l),k}}
where $a_{k,j}$ are anticommuting Grassmann variables,
\eqn\supexpl{r_j\geq 0;\quad r_0=0;\quad \alpha (k)=1+\sum\limits_
{m=1}^{k-1}r_m;\quad \beta (k)=\alpha (k)+r_k-1.}
The symbol $\sum\limits_{\sigma}$ in the \super\ denotes the sum over
all
the
permutations of numbers 1,...,N such that
\eqn\permutrestr{\sigma(\alpha (k))<\ldots <\sigma(\beta (k))\qquad
k=1,\ldots ,M.}
The simplest way to proof \super\ is to perform an induction in M.

\medskip\noindent
{\bf Lemma 2}\smallskip
Consider a $N\times N$ matrix $A$ with elements $a_{ik}$. Suppose we
have a partition of $N$ into $M$ integers: $N=r_1+r_2+\ldots+r_M$ where
$r_j\geq 0$. For each permutation $\sigma$ satisfying the condition
\permutrestr\ we define matrices:
\eqn\kasigma{A_{\sigma}^{(k)}=\left(\matrix{
a_{\sigma (\alpha (k)),\alpha (k)}&\ldots&a_{\sigma(\alpha (k)),\beta
(k)}\cr
\vdots&\ddots&\vdots\cr
a_{\sigma (\beta (k)),\alpha (k)}&\ldots&a_{\sigma (\beta (k)),\beta
(k)}
\cr}\right)}
where $\alpha (k)$ and $\beta (k)$ are defined in \supexpl.
Then
\eqn\determinant{{\rm det}A=\sum\limits_\sigma {\rm sign}(\sigma)
\prod_{j=1}^M {\rm det}|A^{(j)}| }
where $\sigma$ is an arbitrary permutation satisfying
\permutrestr .
This is a generalization of the usual way of computing determinants
(see for example \ref\nl{F. Nei\ss, H. Liermann, Determinanten und Matrizen,
8$^{\rm th}$ ediction, Springer Verlag, (in german).}).
\smallskip\noindent
\smallskip\noindent
The best known subcase of eq. \determinant\
is that for which $r_1=1$, $r_2=N-1$ and remaining $r_j$ vanish.

With the help of Lemmas 1 and 2 we can  check eq. \proptwo. Our aim
is to calculate the correlator $\langle 0|\prod\limits_{I=1}^{N_b}
b(z_I)|0\rangle$.
The strategy is to decompose the fields $b(z_I)$ according to
\splitting .
We apply now the Lemma 1 and
obtain a sum over all possible partitions of the number $N_b$. According
to properties of the vacuum expressed in \vaccond\ only one partition
gives a nonzero contribution: the number of $b_k$ fields has to be
equal $N_{b_k}$. Once we establish this fact we arrive
immediately at the product of $n$ correlators of the type given in eq.
\propone. With the help of Lemma 2 we immediately complete the proof of
the Proposition 2.

\vskip 1cm
\listrefs
\bye